\documentclass[a4paper,12pt]{article}

\begin{document}
\vspace{5 pt}

\centerline{\large \bf On the physical nature of the photon mixing processes}
\centerline{\large\bf in nonlinear optics.}

\vspace{7 pt}
\centerline{\sl V.A.Kuz'menko}
\vspace{5 pt}
\centerline{\small \it Troitsk Institute for Innovation and Fusion Research,}
\centerline{\small \it Troitsk, Moscow region, 142190, Russian Federation.}
\centerline{\small \ E-mail: kuzmenko@triniti.ru}

\vspace{5 pt}
\begin{abstract}

The physical nature of the photon mixing and photon echo processes is 
discussed on the basis of inequality of the forward and backward optical 
transitions.

\vspace{5 pt}
{PACS number: 42.50.-p}
\end{abstract}

\vspace{12 pt}

Wave (photon) mixing processes of different kind play an important role in 
nonlinear optics. Interaction of light with substance can be described both 
from the wave point of view and from that of the quantum insight into the 
nature of light. It has happen historically, that the so-called semiclassical 
theory has got the greatest development [1,2]. The fact of quantum absorption 
and emission of light is recognized here, but all description of the dynamics 
of optical transitions is performed from the pure wave position of the 
so-called rotating wave model. 

The basic problems in this theory arise when one attempts to explain the 
physical nature of the processes underlying the observable effects. Most 
brightly, this difficulty manifests itself in the case of the population 
transfer effect at sweeping of resonance conditions [3]. The theory well 
describes the dynamics of this effect, but fails to explain its physical 
nature. A similar situation exists with a photon echo effect in the gas 
phase. The theory very well explains the principle of a photon echo, and 
well describes its dynamics, but can not explain the specific physical 
mechanism of the process [1]. An attempt to draw the Doppler effect to this 
explanation demonstrates only the inconsistency of the Bloch model in 
explanation of the particular nature of the photon echo. 

In the framework of wave approximation, the reason of laser stimulated wave 
mixing processes is generally believed to be due to the so-called nonlinear 
susceptibility. If the corresponding factor of nonlinear susceptibility is 
great enough, the wave mixing can occur. If this factor is equal to zero, 
the wave mixing is absent. Also, a restriction on the lowest order wave 
mixing process exists, that is connected with the symmetry of the process. 
Namely, the factor of nonlinear susceptibility ${\bf\chi^2}$ is nonzero only 
in a medium without center of inversion [4].

The purpose of the present report is to discuss an opportunity of photon 
mixing explanation from the position of pure quantum approximation. A good 
theoretical basis exists for this approach, as the Dirac equation predicts 
time invariance violation in electromagnetic interactions [5]. For many 
years, however, a generally accepted opinion exists, that electromagnetic 
interactions proceed with T-invariance preservation [6]. Obviously, this is 
a mistake. This point of view does not have any experimental proof. On the 
contrary, the opposite point of view (here we mean the T-invariance 
violation) has a direct and complete experimental proof for the case of the 
photon interaction with molecules [7]. The experiments show, that although 
the integral cross-sections of absorption and stimulated emission of photons 
by atoms and molecules are, obviously, identical (the Einstein coefficients 
are equal), the difference in spectral widths and cross-sections of the 
forward and backward optical transitions can reach several orders of 
magnitude. The principle of inequality of the forward and backward optical 
transitions is quite sufficient for explanation of most nonlinear effects 
from the pure quantum point of view without using any wave approximation.

In the quantum approach, the physical reason of efficient photon mixing is 
probable ultra high cross-section of the backward optical transition to the 
initial state. The "initial state" concept must include, probably, not only 
a set of quantum numbers, but also the orientation of molecules in space and 
the phase of vibration motion. Fig.~1a shows a general four photon mixing 
scheme. Three laser beams, adjusted in the resonance with $0 \rightarrow 1$, 
$1\rightarrow 2$ and $2 \rightarrow 3$ transitions, interact with molecules. 
As a result of the population transfer, directed superradiation on the 
$3 \rightarrow 0$ transition appears, which transfers molecules precisely to 
the initial state. A little more complex scheme is submitted in Fig.~1b. 
Here the directed superradiation arises at first on the $2 \rightarrow 3$ 
transition, for which level 3 is an initial state. Then directed 
superradiation on the $3 \rightarrow 0$ transition should appear, for which 
0 level is an initial state. In this case we have a combined variant of four 
and six photon mixing.

The selection rules can be deduced from the existence of the photon spin. It 
is impossible to return a molecule precisely to the initial quantum state 
using odd number of photons. On the contrary, even number of photons allows 
this to be done. Therefore, only four, six, and eight photon mixing processes 
are possible in gases and liquids. In the solid state, the rotation of 
molecules is absent. A crystal lattice allows to eliminate the problem of 
the photon's spin. Therefore, in the crystals the three- and five- photon 
mixing processes are also observed.

What type of approximation (the wave or the quantum one) is more acceptable 
for description of the photon (wave) mixing effects in nonlinear optics? 
From our point of view, the quantum approximation is more preferable, 
especially in the cases of using the short delayed pulses of laser 
radiation. Temporary delay between pulses badly coordinates to the 
principles of nonlinear susceptibility and wave mixing. On the contrary, 
quantum approximation well admits using the delayed laser pulses. It only 
imposes certain restrictions on the sequence of interactions of molecules 
with laser pulses. Population of quantum levels must be transferred 
consistently and the whole mixing process should return molecules to the 
initial state. The delay between laser pulses allows to study dynamics of 
the ground and various excited states of molecules [8,9]. 

There are many experiments in the literature, which are rather similar in 
arrangement to that shown in Fig.~1, but simpler in implementation. Sometimes 
these experiments are related to the concept of the photon echo. The physical 
mechanism of the classical photon echo in the gas phase can be shown to be a 
result of degenerative four photon mixing in a three-level system, if one 
takes the photon's spin into consideration. The photon's spin can play a 
role of the "Maxwell's demon" [1]. The basic scheme of this process is 
given in the Fig.~2. The photon (1) of the first laser pulse excites the 
molecule and passes to it a rotation moment, connected with the photon's 
spin. Now a process of rotation dephasing of molecules, connected with 
heterogeneity of rotational spectrum, begins. This process has been studied 
in the beautiful experimental work [10]. (It is necessary to note, that the 
probe pulse in this work stimulated a backward Raman optical transition). 
The inhomogeneity of the absorption spectrum is connected with the hyperfine 
splitting due to higher order interactions like the centrifugal distortion, 
the electronic spin-spin splitting, and others. For typical time of the 
photon echo in the gas phase $\sim 1\mu s $ [11] it makes the rotational 
dephasing of molecules quite significant.

At time $\tau $ the second laser pulse starts. Two photons (2 and 3) are 
connected with the second laser pulse (for the case of the two-pulse echo). 
The second photon deexcites molecule and also compensates the rotation moment 
of the first photon. The absorption of the third photon is accompanied by 
transfer of the rotation moment to the opposite direction. After that the 
process of rotation rephasing begins, which is finished to the 
time $ 2 \tau $ by directed superradiation of the photon echo pulse 
(photon 4), which returns the molecule precisely to the initial state. 
Existence of the so-called line wings [12] in the absorption spectrum of 
polyatomic molecules (like $SF_6$ and $BCl_3$) allows to eliminate the 
problem of exact resonance of narrow laser radiation with absorption lines 
of molecules. Thus, inhomogeneity of the absorption spectrum, connected 
with the Doppler effect, has no relation neither to the rotation rephasing, 
nor to the photon echo effect. This simple quantum model predicts also, 
that the main phase-match direction of the photon echo superradiation 
(${\bf k_e = k_1 - k_2 + k_2 = k_1}$) must be collinear with the beam of the 
first pulse. On the contrast, the phase-match direction 
(${\bf k_e = - k_1 + 2k_2 }$) is impossible for the real photon echo, since 
the sign "minus" of vector ${\bf k_1}$ corresponds to stimulated emission of a 
photon.

If the second laser pulse split on two pulses, we obtain the so-called 
stimulated photon echo variant [13,14]. Here a photon of the first laser 
pulse excites molecules and gives a rise to the rotation dephasing process. 
The second laser pulse transfers molecules to the ground state and stops 
the rotation dephasing process. The delay between the second and the third 
laser pulses allows to study the dynamics of the ground state. The third 
laser pulse again excites the molecules and initiates the process of their 
rotation rephasing, which is finished by directed superradiation of a photon 
echo pulse. In this case, the delay between the third laser pulse and the 
photon echo pulse is equal to that between the first two laser pulses. 

The experimental studies of the dynamics of molecules in the liquid phase 
[15-18] are usually associated with stimulated photon echo. Such an 
association for these experiments is not successful. Photon echo is only a 
special case of the four-photon mixing. The main reason of the photon echo 
is laser-induced rephasing process of the rotation motion. If the dephasing 
is not so great, there is no need neither in the rotation rephasing, nor in 
the photon echo. In the general case, only rotational and vibrational 
alignment of molecules is required. In the gas phase, the rotational 
alignment of molecules occurs by their free rotation [10]. In the liquid 
phase the librational motion plays the role of rotation. Therefore the 
duration of a superradiation pulse in four photon mixing in the liquids 
characterizes mainly the period of librational alignment of molecules (if 
the lifetime of excited molecules is sufficiently long). 

The so-called peak shift is measured in these works as one of the most 
important experimental parameters. This peak shift is related to some abstract 
correlation function [15]. However, the effect of peak shift can be given an 
alternative pure physical explanation. When the third laser pulse coincides in 
time with the second one, the shift of pulses characterizes, probably, most 
optimal conditions for population transfer in the system, which leads to 
appearance of the directed superradiation. Here we have a rather general and 
interesting example of the so-called effect of a counterintuitive sequence of 
interactions of molecules with laser pulses. The most effective 
superradiation (the population transfer) takes place when radiation of 
the first (in time) laser pulse interacts with molecules exited by 
radiation of the second laser pulse (some overlapping of laser pulses 
should, certainly, take place). This effect was described long ago in the 
works on the dynamics of population transfer in atoms and molecules in the 
gas phase under influence of nanosecond pulses of laser radiation [19,20]. 
For the photon mixing process of this kind the phase-matched direction for 
superradiation (${\bf k = k_2 - k_1 + k_2 = 2k_2 - k_1 }$) can well be 
realized. It follows from the notes above, that the maximal value of the 
measured peak shift is determined mainly by the shape and width of the used 
laser pulses. 

The delay of the third pulse destroys optimum conditions for the population 
transfer and results in the sharp and substantial reduction in intensity of 
superradiation and measured value of the peak shift in liquid [15]. Probably, 
this is a consequence of librational and vibrational dephasing. The 
temperature dependence of the peak shift in solid samples is especially 
interesting here [21]. May be this dependence characterizes some specific 
characteristics of librational motion.

The optimal experimental conditions are, obviously, different for residual 
superradiation. A dephasing degree of molecules, which are prepared in the 
ground state by the first two laser pulses, can be more substantial. This 
dephasing can be due to the fact, that the molecules stay different time in 
the exited state. The inertia moments of molecules in the ground and excited 
states can be essentially different (especially in the case of electronic 
excitation). The least dephasing in the ground state occurs when the 
molecules spend minimal and equal time in the excited state. This 
condition can be implemented, when the first and the second pulses 
practically coincide in time.

So, from the given point of view the dependence of the peak shift on the 
delay of the third laser pulse, and on the temperature, as well as the 
shape of the superradiation pulse, characterize mainly the dynamics of 
the libration motion of molecules. Experiments, similar to work [10], 
but conducted in a liquid phase, could also give important information on 
the dynamics of librational alignment. The possible role of rotation 
rephasing and the existence of a real stimulated photon echo in the liquid 
phase require further study and discussion. 

The principle of inequality of the forward and backward optical transitions 
is suitable not only for explanation of the physical nature of photon mixing 
and photon echo effects. It allows to explain easily and in the natural way 
such effects, as population transfer at sweeping the resonance conditions 
[3], amplification without inversion [22], coherent population trapping [23], 
electromagnetically induced transparency [24], and others.

Thus, two approaches can be considered for description of the dynamics of 
optical transitions, the wave approximation and the quantum one. The wave 
approximation has a 50-year's old history [2], advanced mathematical tools, 
and some problems with physical interpretation. The quantum approximation has 
a good theoretical base (the Dirac equation), simple and clear physical 
sense, and sufficient proofs. The quantum approximation requires experimental 
study of the basic parameters of the backward optical transitions and 
creation of the mathematical tools for description of the effects in 
nonlinear optics.

\end{document}